\pdfoutput=1
\documentclass[sigconf, nonacm, balance=false]{acmart}
\usepackage{tabularx}

\begin{document}

\title{Does the Citation Matter? Quality, Grounding, and Physician-Rubric Accuracy of Google’s AI Overviews for Medical Information}

\author{Sam Agnoli}
\affiliation{%
  \institution{Northwestern University}
  \city{Evanston}
  \state{Illinois}
  \country{USA}}

\author{Brigid Reilly}
\affiliation{%
  \institution{Northwestern University}
  \city{Evanston}
  \state{Illinois}
  \country{USA}}

\author{Lily Barna}
\affiliation{%
  \institution{Northwestern University}
  \city{Evanston}
  \state{Illinois}
  \country{USA}}

\author{Em{\H{o}}ke-{\'A}gnes Horv{\'a}t}
\affiliation{%
  \institution{Northwestern University}
  \city{Evanston}
  \state{Illinois}
  \country{USA}}

\begin{abstract}
Google’s AI Overviews (AIOs) are the first result users encounter when searching the web, shifting the work of selecting and evaluating sources from the user to Gemini. This may be especially consequential for health information, given that Google’s traditional algorithm was designed to prioritize quality sources, and inaccurate information carries health risks. We evaluated 3,611 benchmark health questions, comparing each AIO against that question’s top five traditional results on source selection and quality. We also evaluated the AIOs' statement grounding in their cited sources and how accurate AIOs were against physician-written rubrics. Overviews appeared for 96.4\% of questions but drew on substantially different evidence than traditional search: only 44.4\% of the top five results were cited by the corresponding AIO, which instead over-represented commercial health publishers (e.g., WebMD, Healthline) at the expense of academic and government sources (e.g., Cleveland Clinic, National Health Service). AIO-cited pages scored lower than traditional results on four of six well-established credibility criteria. The grounding of AIO statements to their sources was poor: after correcting for classifier error, an estimated 56.7\% of cited statements were supported by their cited page, and a further 20.8\% of all statements were factual claims with no citation at all. When AIOs were graded against physicians’ rubrics, they earned 38.2\% of all available points (e.g., accuracy 54.2\%, completeness 21.6\%), and among the subset of AIOs whose physician rubric anticipated a specific inaccurate or potentially harmful claim, 22.2\% of AIOs contained the harmful content. Notably, the accuracy of AIOs was unrelated to the grounding of their statements. We recommend AIO literacy interventions for users and holding AIOs at least to Google’s own health-ranking standards.
\end{abstract}

\keywords{AI Overviews, generative search, health information, search engine audit, retrieval-augmented generation, web search}

\maketitle

\section{Introduction}

Google’s AI Overviews (AIOs) are now the first result its 2.5 billion monthly users encounter when searching for information online~\cite{pichai2026}. In practice, AIOs employ the large language model (LLM) Gemini with retrieval-augmented generation (RAG) to provide tailored answers with citations to users’ questions. AIOs supersede the traditional search algorithm on an opt-out model~\cite{bianchi2025}, functionally shifting the work of selecting and evaluating sources from the user to Gemini. Rather than decomposing a complex question into multiple searches, users can now search for a single question with many details and caveats, and are provided with a synthesized answer from several sources~\cite{google2024}. When users receive an AIO, they are less likely to click links from the traditional search and rarely check the AIO against its sources~\cite{chapekis2025}. This suggests that AIOs’ accuracy and fidelity to their citations are increasingly the primary safeguards of online information for users. However, AIOs were deployed with little understanding of how they might impact the information ecosystem.

Consider that Google’s traditional search algorithm has been carefully developed over the past 30 years to provide high-quality information in a variety of contexts~\cite{google2019a,google2019b}. Google has particularly strict criteria for health information, ranking credible sources like academic and government domains more highly than commercial domains~\cite{google2019b,strzelecki2020}. Yet emerging evidence suggests that AIOs may draw upon lower quality sources than traditional searches. One study of AIOs in non-healthcare domains showed that AIOs rely on more user-generated content (e.g., anonymous social media posts) than traditional search results~\cite{hagar2026}. These lower quality sources could be associated with harmful or misleading claims. For example, an audit of AIOs for baby care and pregnancy questions observed that basic medical safeguards were omitted in the vast majority of AIOs~\cite{hu2026}. A related concern is that AIOs may make claims that are not grounded in their sources, with one study of non-medical queries finding that 11\% of claims were unsupported by their citations~\cite{xu2026}. It is unclear whether AIOs inherit Google’s previous quality standards in health domains, which were developed to protect users from unnecessarily complex and potentially harmful information~\citep[]{kitchens2014,google2019a,google2019b}.

Studying AIOs in the context of healthcare is particularly important given that health-related queries are frequently high-stakes, and inaccurate or misleading information can lead to worse outcomes, including increased mortality~\cite{swirethompson2020}. Related work on health content from state-of-the-art LLMs found that nearly half of common health-related answers are misleading and/or incomplete~\cite{tiller2026}, which is especially concerning given that people trust LLMs as much as doctors~\cite{shekar2025}. A recent systematic review suggested that, in addition to increasing the volume of health misinformation, AI also increases its perceived credibility as compared to human-written content~\cite{saeidnia2026}. Additionally, AIOs~\cite{sun2024} and LLMs~\cite{xu2023} are generally preferred to traditional searches for the rapidity and clarity with which users receive an answer. Taken together, AIOs may represent an outsized threat to the healthcare informational ecosystem that has yet to be characterized.

The current study audited AIOs in the high-stakes domain of health to test whether their citations represent credible sources, grounded statements, and accurate answers. We queried 3,611 health questions to Google from HealthBench~\cite{arora2025}, OpenAI's benchmark for realistic health conversations, each paired with physician-written grading rubrics, and evaluated each AIO in four complementary analyses. First, we examined the degree of overlap between AIOs' sources and the top 5 traditional sources. Second, we compared the quality of sources across both groups using two widely used healthcare website evaluation instruments: Journal of the American Medical Association (JAMA) criteria and Health on the Net Code (HONcode). Third, we examined the degree to which statements within each AIO were grounded by their citations using HallOumi, an open-source claim verification classifier that has previously been used to evaluate AIOs in a similar context~\cite{mickle2026}. Finally, we graded each AIO on physician-written criteria using GPT-4.1 to assess the accuracy and completeness of AIO answers~\cite{arora2025}, and tested whether the quality and grounding of AIOs’ citations predicted their accuracy and completeness.

\subsection{Research Questions and Related Work}

Our first research question was to confirm that AIOs are frequently generated for health queries, even when queries are high in complexity and risk, and whether these AIOs overlapped with traditional search results. Empirical work demonstrated that AIOs are generated more often for long queries as compared to short queries~\cite{grossman2026}, suggesting that they may increase as a function of a query’s complexity. Research has not yet examined whether AIOs are more frequent for high-risk health questions where inaccurate information would be especially deleterious. Once AIOs are created for healthcare questions, do they rely on the same sources as Google’s traditional algorithm? In non-health domains, AIOs reference a more diverse range of sources than is typical of traditional searches~\cite{kirsten2025}. These sources include lower-ranked results that do not appear on the first page of a search and would otherwise have been ignored by most people: studies estimate that users evaluate as few as the first three to five results~\cite{barilan2009,hotchkiss2005}. Finally, we sought to characterize the instances where AIOs generate answers without citing any sources at all.

Our second research question was whether AIOs' sources differed in quality from traditional sources. Though AIOs cite more sources than a user may encounter with a traditional search, this diversity could let in lower quality sources. For example, AIOs frequently index anonymous, uncredentialed user-generated comments on internet forums as evidence for their summaries~\cite{stox2025,hagar2026}. Additionally, traditional ranked searches have various quality assurances developed over years of optimization research~\cite{qin2012,strzelecki2020}, which would ideally be inherited by AIOs.

Our third research question was whether AIOs' statements were grounded in their corresponding citations. LLMs are prone to errors including hallucinations~\cite{belem2025}, though these can be curtailed when LLMs ground their statements on other information through RAG~\cite{lewis2020,mialon2023}. Contemporary LLMs do a decent job of correctly answering general health questions, but performance decreases as queries become more complex~\cite{narula2023,fernandezpichel2025}. Unfortunately, LLMs also tend to be overconfident in their claims~\cite{chen2023,mielke2022}, and LLM citations signal credibility~\cite{ding2025} that could misinform people about the quality of advice~\cite{steyvers2026}. The fidelity of AIOs’ claims to their source is not a given and warrants scrutiny.

Our fourth research question was whether AIOs' responses met physician standards of accuracy and completeness. HealthBench paired each question with its own unique physician-written criteria (\textit{N} = 41,180 criteria) that specified what a good response should include (positive points, $\sim$70\% of criteria) alongside inaccurate or misleading claims (negative points, $\sim$30\% of criteria). Whereas our first three questions provide indirect markers of quality, the physician criteria anchored our evaluation in clinical expertise. We also examined whether accuracy and completeness were related to the other research questions, as all measures were intended to provide complementary evaluations of AIOs for health information.

\section{Methods}

\subsection{AI Overview Questions}

Questions were drawn from OpenAI’s HealthBench~\cite{arora2025}, an open benchmark for AI in healthcare contexts, because its diverse questions may better represent the longer, more conversational queries users now pose to Google after the advent of AIOs~\cite{budaraju2025}. HealthBench is comprised of 5,000 realistic health questions, each with physician-written rubric criteria (e.g., ``States that antibiotics do not treat viral infections,'' +7 points; ``Recommends a double-dose of warfarin after missing a dose,'' $-$8 points) and one of seven themes defined by HealthBench to assess model performance on real-world health interactions. 

We used HealthBench's rubric and themes to classify rubric complexity (low, medium, high) and risk level (low, medium, high), see Table~\ref{table1}. These classifications were included to ease interpretability by conveying 1) how clinically demanding each question is and 2) the level of potential harm that could result if a question is answered poorly, respectively. Rubric complexity was derived from counting rubric criteria as a proxy for how much clinical knowledge an answer demands: low (2--3 rubric criteria), medium (4--7 criteria), and high (8+ criteria). Risk level captures the potential harm of a question when answered poorly or inaccurately. Because HealthBench does not rate risk, we grouped its seven themes by the likely consequence of a poor answer, following the classification in \citet{arora2025}. High risk comprised emergency referrals, where a missed emergency can delay urgent care, and response depth, where an incomplete or oversimplified answer can omit crucial information that a user needs to make a health decision. Medium risk comprised responding under uncertainty and communication adjusted to the user's level of expertise, where the main danger lies in how information is framed (e.g., false confidence about evidence lacking consensus, or explanations provided at the wrong level of expertise). Low risk comprised context seeking (e.g., incomplete questions where the AIO notices and asks for additional details), global health (e.g., tailored advice to local guidelines), and health data tasks (e.g., documentation for a patient to share with their doctor). In these cases, a poor answer was often generic, incomplete, or ill-suited to the setting rather than immediately dangerous. Because this grouping is our own, results for each theme are reported separately (Table~\ref{tab:app-themes}).

\begin{table}[tbp]
  \caption{HealthBench Questions by Complexity and Risk}\label{table1}
  \small
  \begin{tabularx}{\columnwidth}{@{}>{\raggedright\arraybackslash\hsize=1.20\hsize}X>{\raggedleft\arraybackslash\hsize=0.95\hsize}X>{\raggedleft\arraybackslash\hsize=0.95\hsize}X>{\raggedleft\arraybackslash\hsize=0.95\hsize}X>{\raggedleft\arraybackslash\hsize=0.95\hsize}X@{}}
    \toprule
    Classification & Low complexity & Medium complexity & High complexity & \textit{Total} \\
    \midrule
    Low risk & 29 & 148 & 1,102 & 1,279 \\
    Medium risk & 207 & 226 & 1,179 & 1,612 \\
    High risk & 143 & 24 & 553 & 720 \\
    \textit{Total} & 379 & 398 & 2,834 & 3,611 \\
    \bottomrule
  \end{tabularx}
\end{table}

\subsection{Question Selection}

HealthBench included 5,000 realistic health conversations, which we filtered down to 3,611 questions. To be included, questions had to be in English and focused on healthcare problems. Problem-focused questions were feasibly answerable by a medical professional. Common exclusions were conversations that focused on an administrative task (e.g., writing up a patient note) or being too unspecific to answer. These questions varied in word length and were right-skewed (\textit{M} = 43.6, \textit{SD} = 44.5, \textit{Mdn} = 31, range 2--899; Figure~\ref{fig:app-length}).

\subsection{Source Retrieval}

Questions were sent to Google through the SerpAPI endpoint between September 16--18, 2026, with locale parameters fixed to U.S. English searches. Questions were submitted verbatim from HealthBench, including multi-sentence and multi-turn items. One question failed to be retrieved after two tries, leaving 3,610 questions with captured results.

For each question, we captured 1) the AI Overview, including its full text and every source Google cited, and 2) the first five traditional results, excluding video results. Only the first five traditional results were retrieved based on past research suggesting that users read down to around the fifth source~\cite{petrescu2014,abualsaud2019}. For a complete description of scraping details, see Appendices~\ref{app:scrape} and~\ref{app:retrieval}.

\subsection{Source Quality}

To measure the quality of domains AIOs relied on, an automated rule-based Python scorer coded all AIO and traditional sources using the JAMA and HONcode criteria. The JAMA criteria~\cite{silberg1997} is a widely used evaluation tool that assesses four key dimensions: authorship, attribution, currency, and disclosure. Each dimension is rated on a binary scale (i.e., either 0 or 1). To get a perfect score, a source must, 1) provide credentials for all authors, contributors, and/or institutions, 2) disclose whether there are conflicts of interest including funding, sponsorship, advertising, and support, 3) adequately reference all content with sources and attribute copyright data, and 4) provide dates of both initial content and any updates.

We paired JAMA with HONcode, a set of criteria originally designed to certify health websites as trustworthy~\cite{bayani2023}. In the current study, we operationalized HONcode as two binary criteria rather than four: Authority (a credentialed author or reviewer is identifiable, or the publisher is a government or academic institution: .gov, .edu, NHS, CDC, NIH, or a major academic health system) and Complementarity (the page positions itself as supporting rather than replacing the clinician relationship), for a total of 0–2. The remaining two criteria were dropped because HONcode Attribution was identical to JAMA Attribution, and Justifiability could not be coded reliably. Coding rules are listed in Table~\ref{tab:app-jama}. 

Automated scoring was compared to a human-coded subset of 3 representative pilot questions (174 judgments, see Table~\ref{tab:app-agree}). The automated scorer agreed with human coders on 84\% of judgments, yielding substantial agreement (Cohen's $\kappa$\textit{ =} .68).

\subsection{Statement Grounding (HallOumi)}

To examine whether statements were grounded to their source, we used the publicly available classifier model HallOumi (oumi-ai/HallOumi-8B-classifier), which was designed to test AI-generated text against source documents. HallOumi returned a per-statement supported/unsupported verdict with a confidence score~\cite{oumi2025}.

This procedure yielded four verdicts. Supported: cited and backed by a cited page. Unsupported: cited, all cited pages readable, and none support the statement. Indeterminate (reported as skipped): the readable cited pages did not support the statement, and at least one cited page could not be recovered, so support could not be ruled out. Uncited: Google attached no citation. Across all overviews, 24,060 statements were identified, of which 12,175 were scorable against a cited source. We reported grounding as the proportion of scorable statements that are supported. Uncited factual statements were reported separately as a frequency. We manually coded a subset of 100 statements for grounding, and HallOumi agreed with human coders on 81\% of judgments, yielding substantial agreement (Cohen's $\kappa$\textit{ =} .61). In 16 of the 19 disagreements, human coders judged the statement supported and HallOumi did not, so HallOumi’s support rates are likely conservative; we report them as lower bounds, alongside a rate corrected for HallOumi’s misclassification~\cite{rogan1978} (see Appendix~\ref{app:halloumi}). For an illustrative example of an unsupported statement (agreed upon by both human coders and HallOumi), a parent asked whether she should call a doctor after a child bumped their head. The AIO instructed her to keep the child awake, despite none of the cited pages providing such instruction.

\subsection{Accuracy Against Physician Rubrics}

To assess the accuracy of AIO statements, we employed GPT-4.1 to grade the AIOs against HealthBench’s physician-written rubric~\cite{arora2025}. Each rubric criterion describes the content required for a correct answer specific to the corresponding question, ranging from –10 to 10 points (with negative points referring to misleading/inaccurate content). Across the usable AIOs, there were 41,180 rubric criteria (median 12 per question), of which 12,218 (\textasciitilde{}30\%) described content to avoid. Grading required semantic judgment rather than keyword matching: an AIO could satisfy each criterion through paraphrase, implication, or partial coverage. Because our coders did not have clinical expertise, and grading all 41,180 AIO–criterion pairs by hand was not feasible, we followed HealthBench’s protocol and used its physician-validated grader. In a random subsample of 3,000 physician-graded examples, we replicated Arora et al.’s finding that GPT-4.1 agreed with physicians (macro-F1 = .707) more closely than physicians agreed with one another (.681)~\cite{arora2025}.

AIOs' scores were the sum of the criteria judged as met by GPT-4.1, divided by the maximum possible total points. Accuracy refers to content that is medically correct, hedging when evidence is uncertain. Completeness refers to physicians’ expectations for responding to a patient, including the key steps, caveats, and symptom red flags to be aware of. We focus analyses on the accuracy and completeness criteria to aid in interpretation (together, the clinical score), however we report results for all of HealthBench’s criteria: context awareness, communication quality, instruction following, consensus criteria, no factually incorrect information (i.e., general consensus criteria), flagged as inaccurate or potentially harmful (i.e., specific penalty criteria distinct from general consensus criteria), and referred appropriately (see \citet{arora2025} for definitions of each criterion).

\subsection{Analysis}

The unit of analysis was the HealthBench question. For each question and measure, the mean score of the AIO's cited pages was paired with the mean score of that question's traditional top five, and the two groups were compared with a paired samples \textit{t}-test. Effect sizes were reported as Cohen's \textit{d}. \textit{p} values were corrected with the Benjamini–Hochberg False Discovery Rate (FDR) procedure across the six JAMA and HONcode quality measures. The same paired comparison was repeated within each risk level and each rubric complexity level. Citation presence (whether an overview carried any references at all) and statement support rates were compared across risk and complexity levels with Cochran–-Armitage tests and chi-square tests, respectively. Physician-written rubric scores were compared across risk and complexity levels and correlated with JAMA, HONcode, and grounding scores using Spearman’s $\rho$. As a robustness check, all main analyses were repeated on single-turn questions of 44 words or fewer (the full-sample mean).

\section{Results}

The analytic sample included 3,481 questions with retrieved AIOs. Of these, 3,474 had usable AIO text (at least 200 characters), and 3,103 also had at least one readable webpage in both the AIO and traditional groups, forming the sample for all paired source comparisons (see Table~\ref{tab:app-sample}; for descriptives, see Table~\ref{table2}). AIOs averaged 283.2 words (\textit{SD} = 121.7). For AIOs that cited at least one source (\textit{n} = 3,154)\textit{,} they cited on average 8.50 sources (\textit{SD} = 4.07). However, some AIOs cited no sources \textit{(}\textit{n} = 327). In contrast, we capped traditional searches to five sources and usually retrieved all 5 sources (\textit{n} = 3,322; four or less sources \textit{n} = 159).

\begin{table}[tbp]
  \caption{Descriptive Statistics for the Analytic Sample. \textit{n} counts questions that returned an AI Overview. Traditional results are capped at the top five by design. AI Overview citations are captured uncapped.}\label{table2}
  \small
  \begin{tabularx}{\columnwidth}{@{}>{\raggedright\arraybackslash\hsize=2.00\hsize}X>{\raggedleft\arraybackslash\hsize=0.75\hsize}X>{\raggedleft\arraybackslash\hsize=0.75\hsize}X>{\raggedleft\arraybackslash\hsize=0.75\hsize}X>{\raggedleft\arraybackslash\hsize=0.75\hsize}X@{}}
    \toprule
    Measure & \textit{n} & \textit{M} & \textit{SD} & Range \\
    \midrule
    AI Overview length (words) & 3,481 & 283.2 & 121.7 & 7–1,983 \\
    Sources per question: all AI Overview & 3,481 & 7.70 & 4.60 & 0–38 \\
    Sources per question: AI Overview, citing overviews only & 3,154 & 8.50 & 4.07 & 1–38 \\
    Sources per question: traditional & 3,481 & 4.84 & 0.81 & 0–5 \\
    \bottomrule
  \end{tabularx}
\end{table}

\begin{figure*}[tbp]
  \centering
  \includegraphics[width=0.40\textwidth]{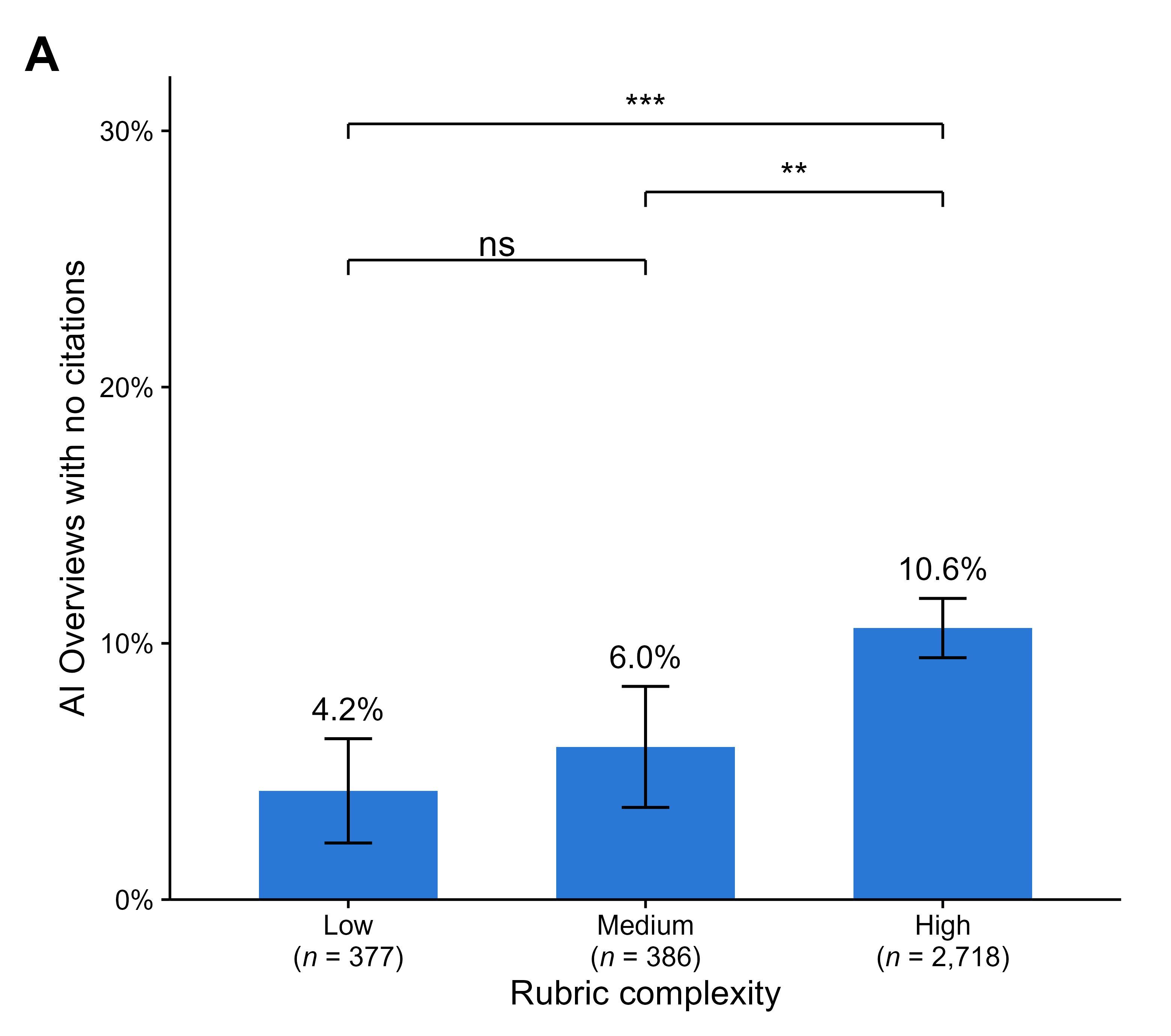}\hspace{0.04\textwidth}%
  \includegraphics[width=0.40\textwidth]{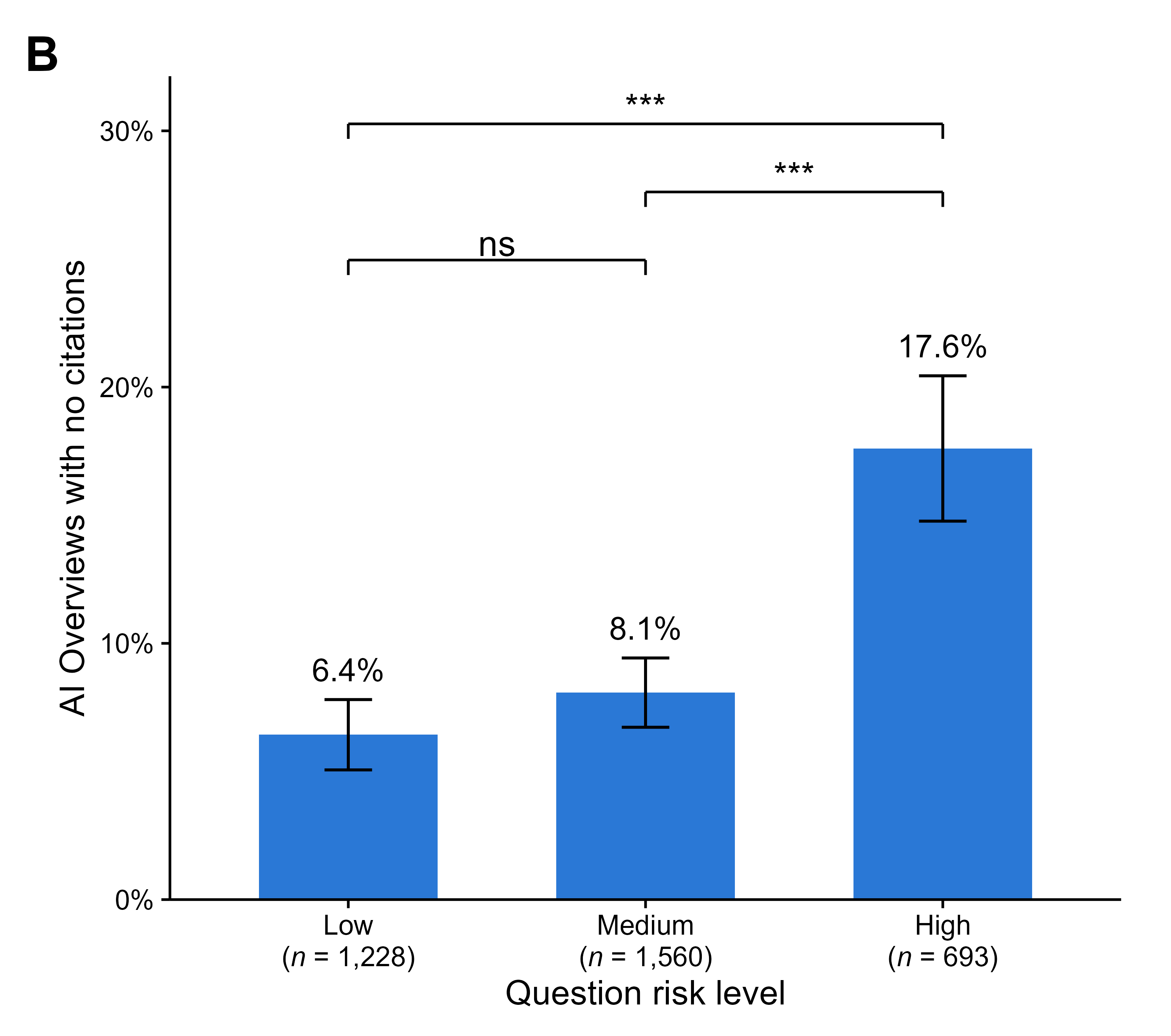}
  \caption{Proportion of AI Overviews Citing No Sources by (A) Rubric Complexity and (B) Question Risk Level. \textit{N} = 3,481 overviews. Error bars are 95\% confidence intervals. Brackets mark pairwise (Holm-adjusted) chi-square tests within each panel. *\textit{p} \textless{} .05. **\textit{p} \textless{} .01. ***\textit{p} \textless{} .001.}\label{figure1}
  \Description{Proportion of AI Overviews Citing No Sources by (A) Rubric Complexity and (B) Question Risk Level}
    \label{fig:fig1}
\end{figure*}

To explore whether the occurrence of citations in AIOs was associated with question characteristics, we examined if question length, complexity, and/or risk differed between cited AIOs and AIOs without citations (see Figure~\ref{fig:fig1}). AIOs without citations had greater question lengths (\textit{M} = 73.5, \textit{SD} = 80.5) than AIOs with citations (\textit{M} = 40.5, \textit{SD} = 38.0), Welch \textit{t}(341.2) = 7.34, \textit{p} \textless{} .001, Cohen’s \textit{d} = 0.75. Similarly, we ran a Cochran--Armitage Chi-squared test to determine if there was an ordinal relationship between AIO citations and complexity or risk level. Uncited AIOs were less frequent for questions with low (4.2\% uncited) and medium complexity (6.0\%) as compared to high complexity (10.6\%), $\chi$\textsuperscript{2}(1) = 20.96, \textit{p} \textless{} .001. For risk level, uncited AIOs were also less frequent for questions with low (6.4\%), and medium risk (8.1\%) as compared to high risk (17.6\%), $\chi$\textsuperscript{2}(1) = 55.57, \textit{p} \textless{} .001. This shows a striking pattern: as questions become longer, more complex, and riskier, AIOs are less likely to cite sources while still generating a response.

\subsection{Overlap Between AI Overview Citations and Traditional Results}

Next, we examined whether AIOs and traditional searches drew upon the same sources across the 31,688 unique question–URL pairs among accessible sources. Because traditional search represents the baseline for Google search and was limited to 5 sources (\textit{n} = 14,443 total sources) compared to the uncapped AIO (\textit{n} = 23,657 total sources), AIO only citations (\textit{n} = 17,245) are less informative than traditional only or shared citations. Across the traditional top 5 results, 44.4\% were also cited by the AIO (\textit{n} = 6,412), compared to 55.6\% (\textit{n} = 8,031) that were only cited by the traditional results. This suggests that Google’s healthcare algorithm substantially diverges from Gemini’s AIOs when gathering relevant sources.

We further characterized search overlap by examining whether the two groups used similar host sites. We found that academic health systems and government health services were more frequently cited by traditional results, and commercial health publishers were cited more by AIOs (see Table~\ref{tab:app-domains}). This provides initial evidence that AIOs draw on less rigorous sources than traditional search results.

\subsection{Source Quality}

\begin{figure}[tbp]
  \centering
  \includegraphics[width=\columnwidth]{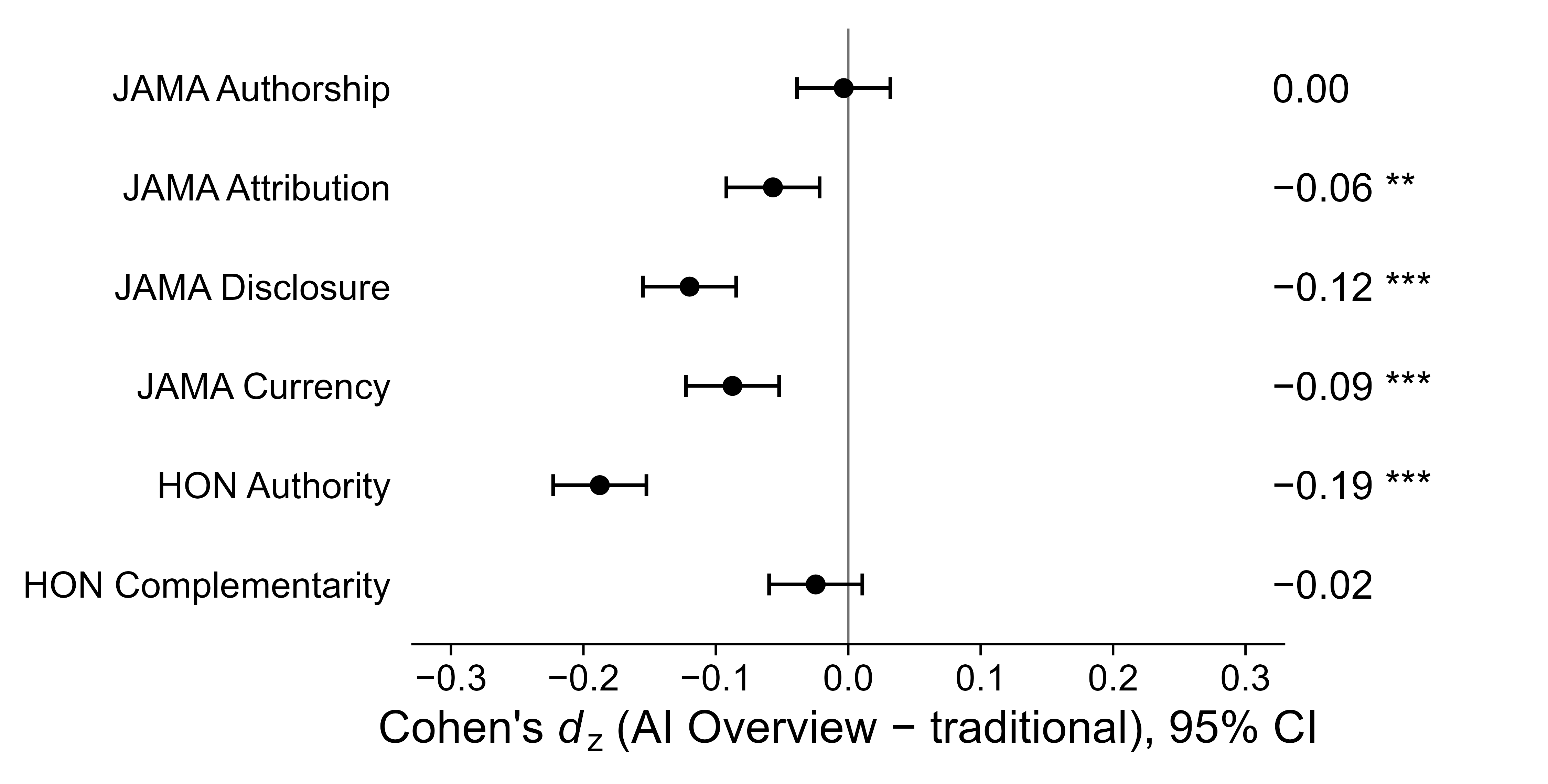}
  \caption{Paired Effect Sizes for JAMA and HONcode Measures. Points show Cohen's \textit{d}\textit{\textsubscript{z}} for AI Overview citations minus traditional results, with 95\% confidence intervals, across 3,103 paired questions. *\textit{p} \textless{} .05. **\textit{p} \textless{} .01. ***\textit{p} \textless{} .001, Benjamini–Hochberg adjusted.}\label{fig:fig3}
  \Description{Paired Effect Sizes for JAMA and HONcode Measures}
\end{figure}

To examine whether AIOs cited the same quality of sources as compared to traditional searches, we coded both following the JAMA and HONcode criteria. After FDR correction, four of six quality measures differed between groups. Traditional sources scored higher than AIO sources on all of those measures. The largest traditional advantage was for HON Authority (\textit{d}\textit{\textsubscript{z}} = $-$0.19, \textit{p} \textless{} .001), followed by JAMA Disclosure (\textit{d}\textit{\textsubscript{z}} = $-$0.12, \textit{p} \textless{} .001), JAMA Currency (\textit{d}\textit{\textsubscript{z}} = $-$0.09, \textit{p} \textless{} .001), and JAMA Attribution (\textit{d}\textit{\textsubscript{z}} = $-$0.06, \textit{p} = .002). JAMA Authorship (\textit{d}\textit{\textsubscript{z}} = 0.00) and HON Complementarity (\textit{d}\textit{\textsubscript{z}} = $-$0.02) did not differ between groups (\textit{p’s} \textgreater{} .205). All measures are presented in Table~\ref{tab:app-quality} and visualized in Figure~\ref{fig:fig3}.

To further explore differences in quality, we repeated the same paired comparison within each risk level and each rubric complexity level, and found that traditional results were favored across all comparisons (Appendix~\ref{app:quality}).

\subsection{Statement Grounding (HallOumi)}

\begin{figure}[tbp]
  \centering
  \includegraphics[width=\columnwidth]{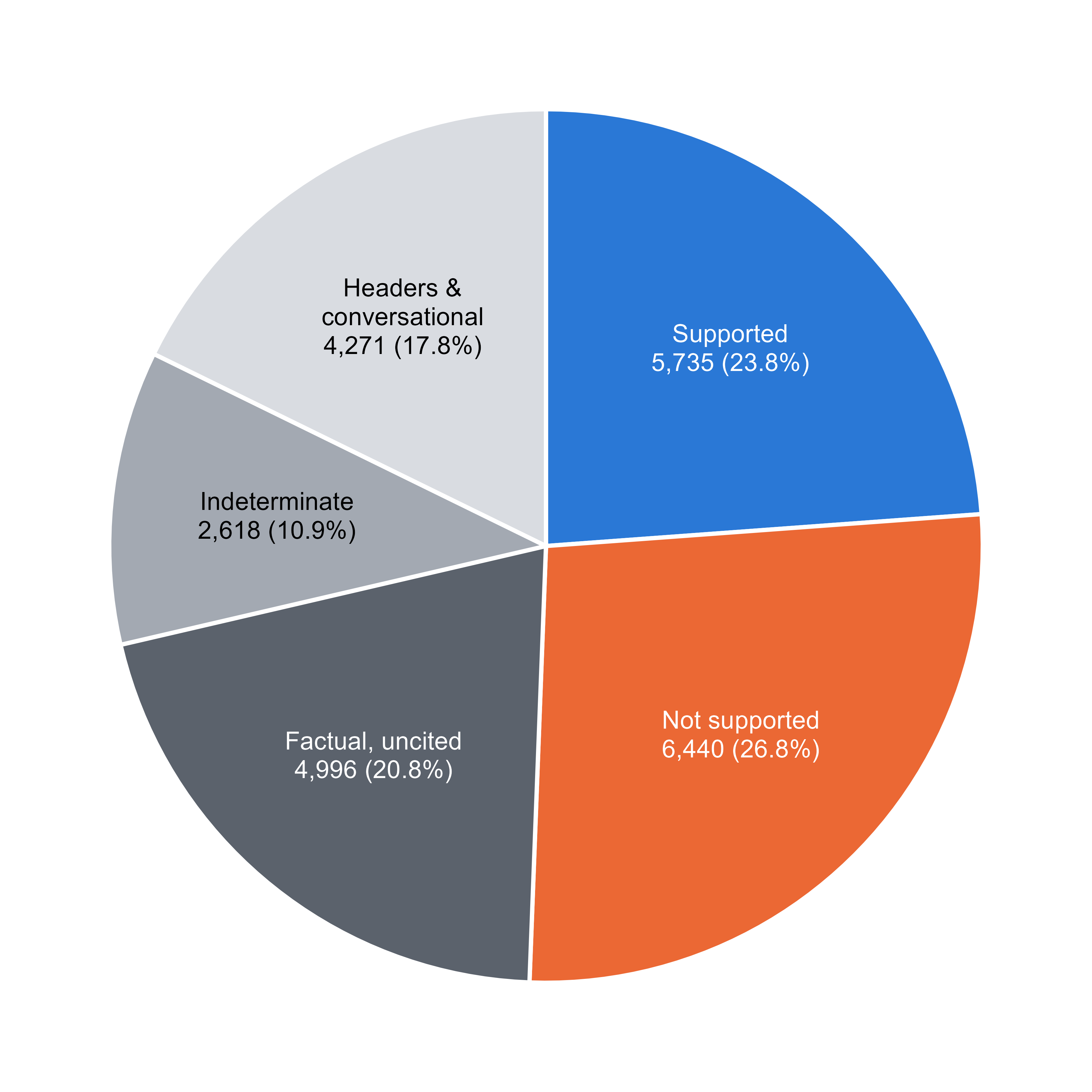}
  \caption{Composition of AI Overview Statements. All 24,060 statements across 3,474 overviews. Only supported and not supported statements (\textit{n} = 12,175) enter the support rate. Indeterminate statements had at least one cited page that could not be read.}
  \label{fig:fig4}
  \Description{Composition of AI Overview Statements}
\end{figure}

The 3,474 AIOs comprised 24,060 statements (see Figure~\ref{fig:fig4}). Only 50.6\% (\textit{n} = 12,175) of the statements could be scored against their AIO citations, with the rest either being uncited factual statements (20.8\%, \textit{n} = 4,996), headers or conversational text (17.8\%, \textit{n} = 4,271), or other indeterminate statements (10.9\%, \textit{n = }2,618). We then tested whether the statements with direct AIO citations were appropriately grounded by deploying HallOumi to judge the probability that each statement was supported by the cited text. Within the directly cited statements, only 47.1\% (\textit{n} = 5,735) of the statements were supported, compared to the 52.9\% (\textit{n} = 6,440) statements that were not supported. Because HallOumi was stricter than human coders, we corrected for its misclassification, estimating that 56.7\% (95\% CI [45.3, 68.6]) of scorable statements were supported. The low support rate, in addition to the sizable chunk of factual statements that had no attached citation (\textit{n} = 4,996), suggests that around half of the cited statements made by AIOs were not adequately grounded.

\begin{figure*}[tbp]
  \centering
  \includegraphics[width=0.40\textwidth]{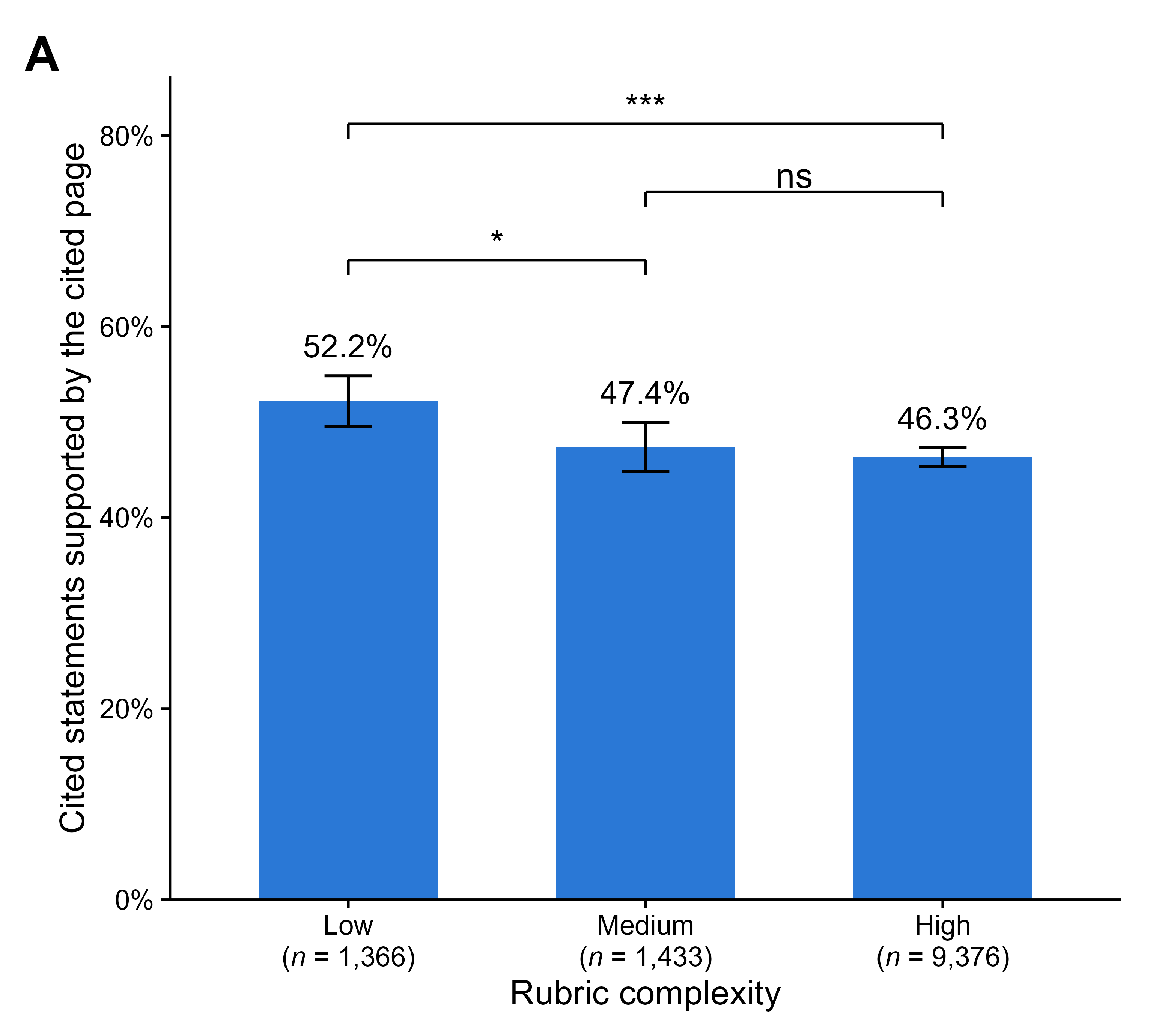}\hspace{0.04\textwidth}%
  \includegraphics[width=0.40\textwidth]{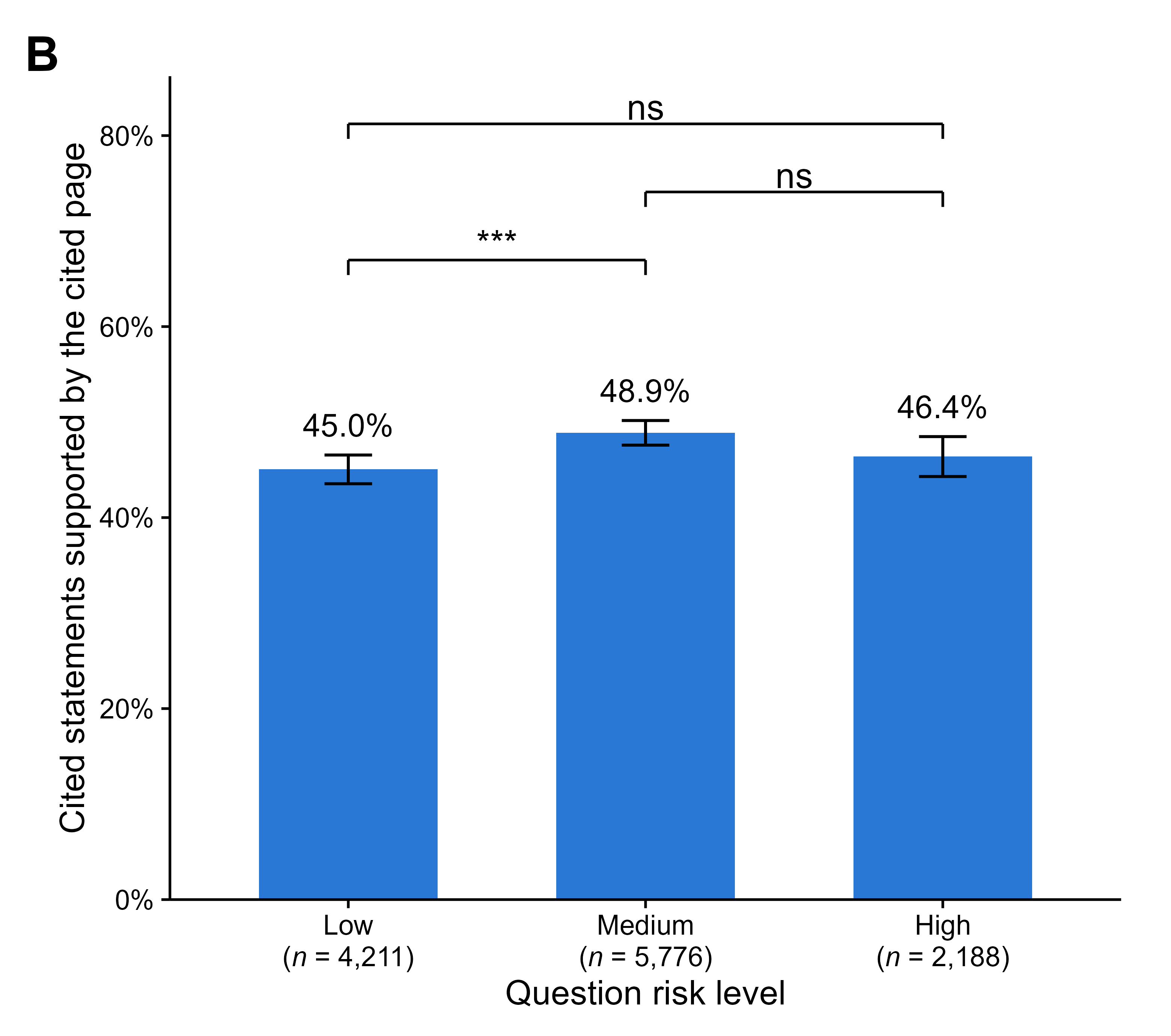}
  \caption{Proportion of Cited Statements Supported by a Cited Page, by (A) Rubric Complexity and (B) Risk Level. \textit{n} = scorable cited statements in each level. Error bars are 95\% confidence intervals. Brackets mark pairwise (Holm-adjusted) chi-square tests within each panel.  *\textit{p} \textless{} .05. **\textit{p} \textless{} .01. ***\textit{p} \textless{} .001.}
  \label{fig:fig5}
  \Description{Proportion of Cited Statements Supported by a Cited Page, by (A) Rubric Complexity and (B) Risk Level}
\end{figure*}

To characterize why some statements were more supported than others, we compared the statements ordinally by complexity and risk level. Support rates differed by rubric complexity, \textit{$\chi$}\textsuperscript{2}(2, 12,175) = 16.57, \textit{p} \textless{} .001, declining from Low (52.2\%) to Medium (47.4\%) to High complexity (46.3\%; see Figure~\ref{fig:fig5}A). Support rates also differed by risk level, \textit{$\chi$}\textsuperscript{2}(2, 12,175) = 14.86, \textit{p} \textless{} .001. However, the pattern was nonlinear, with support lowest for Low-risk questions (45.0\%), highest for Medium-risk questions (48.9\%), and intermediate for High-risk questions (46.4\%); see Figure~\ref{fig:fig5}B.

\subsection{Accuracy Against Physician Rubrics}

AIOs earned 38.2\% of available points across all criteria, notably scoring 39.3\% on accuracy and completeness (54.2\% accuracy; 21.6\% completeness, for all criteria, see Table~\ref{table3}). Most of the accuracy points AIOs lost reflected missing content rather than stated errors (89\% vs. 11\% of points lost, respectively). On HealthBench's general consensus criterion, 93.4\% of AIOs contained no factually incorrect information; however, this general check understates the total errors. For the 2,065 questions where physicians named a specific inaccurate or potentially harmful claim to avoid (e.g., recommending monitoring blood levels every 2 weeks for a drug that could cause toxicity within days), 22.2\% of AIOs made at least one such claim, including 28\% of AIOs that had passed the general check. Additionally, of the 413 emergency-referral questions, AIOs referred to care with appropriate urgency 94.9\% of the time. Taken together, AIOs fell short of physician standards chiefly through errors of omission, but in one in five cases where physicians anticipated a specific and potentially harmful error, the AIO made it.

\begin{table}[tbp]
  \caption{HealthBench Rubric Scores of AI Overviews. Scores are mean percentages of available points (points earned divided by maximum positive points per AIO); per-AIO scores can be negative, and group means are clipped to 0–100\% as in HealthBench. Axis and consensus rows include only AIOs whose rubric had such criteria. Rates include only AIOs whose rubric contained the relevant criterion. \textit{n} = AIOs.}\label{table3}
  \small
  \begin{tabularx}{\columnwidth}{@{}>{\raggedright\arraybackslash\hsize=2.00\hsize}X>{\raggedleft\arraybackslash\hsize=0.50\hsize}X>{\raggedleft\arraybackslash\hsize=0.50\hsize}X@{}}
    \toprule
    Measure & \textit{n} & \textit{\% of available points} \\
    \midrule
    \textit{Score (\% of available points)} &  &  \\
    \quad All criteria & 3,474 & 38.2 \\
    \quad Accuracy and completeness (clinical score) & 3,460 & 39.3 \\
    \quad Accuracy & 3,016 & 54.2 \\
    \quad Completeness & 2,922 & 21.6 \\
    \quad Context awareness & 2,401 & 32.3 \\
    \quad Communication quality & 1,628 & 64.7 \\
    \quad Consensus criteria only & 2,764 & 84.2 \\
    \quad Instruction following & 477 & 32.4 \\
    \textit{Rate (\% of AI Overviews)} &  &  \\
    \quad Contained content flagged as inaccurate or potentially harmful & 2,065 & 22.2 \\
    \quad Met ``no factually incorrect information'' & 861 & 93.4 \\
    \quad Referred appropriately (emergency-referral questions) & 413 & 94.9 \\
    \bottomrule
  \end{tabularx}
\end{table}

We next examined whether the mean of accuracy and completeness scores differed across the level of complexity and risk within a question. Scores were highest for low-complexity questions (76.7\%) as compared to medium (28.8\%) and high complexity questions (35.7\%). This may be explained by the rubrics which included greater consensus for low complexity questions (i.e., 57\% of low complexity accuracy and completeness criteria, vs. 4\% and 9\% for medium and high complexity, respectively), and AIOs met these general consensus criteria far more often than question-specific rubric criteria (91\% vs. 43\%). Clinical scores increased for questions with greater risk levels (low: 26.7\%, medium: 43.4\%, high: 52.4\%). Upon closer inspection, these differences likely reflect the question’s theme (Table~\ref{tab:app-themes}): high-risk questions scored highest because they include the emergency-referral conversations, which earned more points than any other theme. With referral criteria removed, high-risk questions (35.9\%) scored similarly to medium-risk questions (41.3\%) but still higher than low-risk questions (22.5\%).

Next, we examined whether accuracy and completeness differed by JAMA and HONcode quality ratings. Surprisingly, accuracy and completeness were weakly but negatively related to JAMA scores (\textit{$\rho$} = $-$.05, \textit{p} = .003). In contrast, HONcode was positively associated with scores (\textit{$\rho$} = .10, \textit{p} \textless{} .001).

Finally, we examined whether grounding was related to accuracy and completeness scores. Again, counter to our expectations, there was no relationship between grounding and the accuracy and completeness scores (\textit{$\rho$} = .01, \textit{p} = .467; accuracy only \textit{$\rho$} = .01, \textit{p} = .759; completeness only \textit{$\rho$} = $-$.04, \textit{p} = .055). Likewise, AIOs that cited no sources (\textit{n} = 320) scored similarly to cited AIOs (\textit{n} = 3,140) on accuracy and completeness (\textit{M} = 40.4\% vs. 39.2\%), Welch \textit{t}(381.6) = 0.46, \textit{p} = .649. One possibility is that AIOs may be generated primarily by the model weights (i.e., as opposed to RAG relying on the sources for information), which could also be reflected in the poor grounding of statements to their citations.

\subsection{Robustness Check for Short Single-Turn Questions}

One possibility is that the long and/or multi-turn HealthBench questions do not represent the standard search behaviors of users. To address this, we repeated the analyses on single-turn questions of 44 words or fewer (the mean question length, \textit{n} = 1,714). Results did not change the main conclusions (Table~\ref{tab:app-robust}), and effect sizes were similar in all cases with the following exceptions. Uncited AIOs were rarer (from 9.4\% to 5.8\%) and less concentrated in high-risk questions. Also, JAMA Attribution no longer differed between groups, and grounding had a weak correlation with accuracy and completeness ($\rho$ = .06, \textit{p} = .013).

\section{Discussion}

The current study characterizes the state of AIOs in a health context, which may depart from Google’s traditional search in several deleterious ways. AIOs cite more commercial sources that are lower in quality, credibility, and transparency than traditional searches. Further, AIO statements are poorly grounded in their corresponding citations, which worsened in medium and high complexity health questions. AIOs' answers do not comprehensively answer health questions as defined by physicians and also include inaccurate or misleading information one-fifth of the time. These findings have implications for how users might rely on (or avoid) health information presented by AIOs and highlight the need for AI literacy interventions to teach users to verify AI-generated health claims against their sources. We also make recommendations for Google to hold AIOs to the same standards as their traditional algorithm where possible, and to hide AIOs where it is not reliable.

We found that AIOs present largely different sources than traditional results for the same 3,611 health questions. To our surprise, fewer than half (44.4\%) of the traditional top five results were cited by the AIO, despite the fact that our AIOs cited on average 8.5 sources. Further, AIOs occasionally cited nothing despite still generating a response, and this was most common for questions that were longer, more complex, and higher risk. When citations were present, AIOs tended to cite more commercial health publishers (e.g., WebMD, Healthline), in contrast to the academic and government-leaning domains favored by traditional results (e.g., Cleveland Clinic, Centers for Disease Control and Prevention).

Compared to traditional sources, AIOs’ sources scored lower on four out of six JAMA and HONcode website evaluation criteria. The overall difference in effects was relatively small (all $|$\textit{d}\textsubscript{z}$|$ $\leq$ 0.19) and strongest for HONcode’s Authority and JAMA’s Disclosure, suggesting that AIOs relied on domains that were less transparent. This contrasts with Xu et al.’s~\cite{xu2026} findings in non-medical queries that AIO cited domains were more credible than the first page of traditional results in general search queries, though they used a coarser measure that only examined domains rather than page-level criteria. Taken together, these results suggest that AIOs are less strict than Google’s traditional algorithm in providing high-quality health sources.

Statements within AIOs were also poorly cited. Approximately half (50.6\%) of AIO statements were able to be checked, with many factual statements not having corresponding citations. Further, only 47.1\% of statements were supported by their citation, which was noticeably lower than Oumi’s non-academic evaluation of the same HallOumi classifier model (67\% supported) featured in the \textit{New York Times}~\cite{mickle2026}. The unsupported rate also far exceeds the 11\% unsupported rate observed by Xu et al.~\cite{xu2026}, though their questions were much simpler, and used Grok 4.1 Fast Reasoning instead of a classifier model like in the current study. Indeed, support of statements in our study went down as the complexity of the query increased, which might be why support rates were lower than the above findings. We also note that HallOumi was stricter than human coders, who judged 67\% of a 100-statement validation subsample supported, compared with HallOumi’s 54\%. Correcting for this conservativeness raises the sample-wide estimate to 56.7\%. In all cases, the poor support rates are at odds with users' high trust in LLMs (e.g., comparable to doctors~\cite{shekar2025}), particularly in the health domain where misplaced credibility could have dangerous consequences.

Grading AIOs against HealthBench’s physician-written rubrics provided a complementary view of their quality against experts' standards. Our findings suggest that AIOs struggle with errors of omission and inaccurate or misleading statements. Specifically, AIOs earned fewer than two in five of the available points on accuracy and completeness criteria (39.3\%), often losing points in omissions of the specific content physicians expected from each question (completeness: 21.6\%). Encouragingly, AIOs scored slightly higher on accuracy than the LLMs on HealthBench, though these models were \textasciitilde{}2 years older than the current Gemini model~\cite{arora2025}. In contrast, AIOs performed worse than the top models on completeness, likely because AIOs' short answers were penalized as compared to LLMs which had less stringent word limits. AIOs also referred to care with appropriate urgency in the vast majority of emergency-referral conversations (94.9\%), though any omissions in an emergency scenario may be unacceptable. Perhaps most concerning, errors often occurred where physicians had anticipated them: more than one in five AIOs whose rubric anticipated these errors contained content flagged as inaccurate and/or potentially harmful.

Accuracy and completeness scores were also lower for medium- and high-complexity questions than for low-complexity questions, but did not decrease with risk level. The quality of AIOs' sources had a small effect on scores: AIOs scored slightly higher when their cited pages scored higher on HONcode, but slightly lower when their cited pages carried more of JAMA’s transparency markers, which describe how a page is authored and documented. This latter finding may reflect consumer health websites and blogs that frequently disclose authorship, a practice that may have no bearing on the veracity of information.

Notably, the accuracy of AIOs was unrelated to their grounding, and AIOs that cited nothing scored similarly to cited AIOs on physicians' rubrics. Several possibilities could individually or jointly account for this finding. First, the online sources may not be sufficiently up to date, or may not provide consensus healthcare advice, which would limit the association between retrieval and accuracy. Second, AIOs may rely more on Gemini's model weights than on retrieved sources, in which case citations may not contribute meaningfully to accuracy. However, those model weights may have been trained on the same sources as AIOs’ citations. Third, the citations may not reflect what the model drew on. That is, AIOs may first generate a response and then seek out citations that align with the response. Our project cannot distinguish between these explanations, which would require explicitly comparing Gemini's advice without RAG to AIOs with RAG. In any case, this mismatch is particularly concerning given that people trust LLMs more when they include citations than when they do not~\cite{ding2025}. These findings underscore the need for interventions that teach users to carefully check AIOs’ health advice against the sources that they cite.

Because of the size of the dataset, our measures relied on automated coding, the classifier HallOumi model, and GPT-4.1. While these measures all had reasonable agreement with human coders or trained physicians, they nonetheless should not be taken to reflect the absolute quality, grounding, and accuracy of AIOs, even for this subset of healthcare questions. Rather, they should be interpreted as relative comparisons in a health context where providing any information that is inaccurate could potentially prove fatal. We emphasize that even at the upper bounds of our confidence intervals, none of these estimates reached an acceptable standard for health information: fully grounded, accurate, and complete answers.

We hope that this research educates users about the dangers of relying on AIOs when searching for health information. Until AIOs are improved, it may be prudent to hide or ignore AIOs rather than trust them. In this case, citations were not sufficient evidence that AIOs' claims were supported, nor did AIOs' citations improve their accuracy. For Google and other search engines, we strongly recommend that AIOs are held to the same standard as Google’s traditional healthcare algorithms~\cite{strzelecki2020}, perhaps by having Gemini preferentially cite academic and government domains over consumer domains. Further, we urge Google to disable AIOs (or refuse to provide advice and instead refer to a doctor) in cases when questions become too complex or risky. This is an easy step to reduce the situations in which AIOs are more likely to be dangerously misleading.

This was an exploratory first study on a nascent problem, and there are several limitations to consider. The HealthBench questions were developed to benchmark LLM conversations and are therefore longer than typical Google queries. This raised the rate of uncited AIOs, though we note that it may be representative of the longer questions users ask AIOs compared to Google before the advent of AIOs~\cite{budaraju2025}. Additionally, the HealthBench rubric was developed for AI-generated responses that were slightly longer than what AIOs generated, and higher scores in our study were correlated with length \textit{($\rho$} = .14)\textit{,} which may have negatively impacted scoring for the shorter AIOs. We also note that several pages, including those that blocked automated access and social media threads, could not be scraped and so were not included in analyses. Finally, our scrape occurred in a 3-day period, and AIOs are constantly shifting with improvements in the model. It would be beneficial for future work to continue examining AIOs in similar high-impact, controversial topics (e.g., science, politics, conspiracies) where the quality of advice is important for its impact on the informational ecosystem and people's behaviors.

\section{Conclusion}

Google’s AIOs show considerable promise for increasing the efficiency and accessibility of health information. However, our project raises healthcare concerns regarding the quality of AIOs' sources, the fidelity of statements to those sources, and the overall accuracy of AIOs. In doing so, we hope that people seeking health advice will use AIOs with caution (if at all), and that Google and other web directories will expediently address these problems. Billions of people rely on search engines to make informed decisions in their lives. If AI is to be the default tool for these searches, it should supply accurate, grounded information. When that is not possible, AIOs should not mislead us.

\section{Data and Code Availability}

The full pipeline and dataset are available at [\url{https://github.com/SamAgnoli/AIOverviews}]. The repository contains annotated notebooks for each stage of the study and the full dataset is archived on Zenodo [\url{https://doi.org/10.5281/zenodo.23601246}]. We encourage future audits to use the pipeline to collect new data.

\bibliographystyle{ACM-Reference-Format}
\bibliography{references}

\appendix
\emergencystretch=1.5em  

\renewcommand{\thefigure}{A\arabic{figure}}
\renewcommand{\thetable}{A\arabic{table}}
\setcounter{figure}{0}
\setcounter{table}{0}

\section{Methods Details}

\subsection{Search Capture}\label{app:scrape}
Each question was submitted verbatim from HealthBench to Google through SerpAPI’s Google engine, with English and United States selected as the language and location, and requesting the first page of results.

\textit{AIO citations.} All video links (e.g., YouTube, TikTok, and Instagram reels) were excluded because they carry no page text and therefore could not proceed with the automated JAMA and HONcode.

\textit{Traditional results.} The traditional comparison set was the first five non-video results in Google’s traditional ranked list, which does not include the AIO.

\subsection{Page Retrieval and Text Extraction}\label{app:retrieval}
Each page was fetched through a hierarchy of methods, moving to the next method when the previous one failed: (1) a direct HTTP request; (2) the trafilatura fetcher; (3) a headless Chromium browser (Playwright) that executes JavaScript; (4) the page’s AMP version, tried only after a paywall or bot-block response (HTTP 401, 402, or 403, or a bot-check page); (5) the Jina Reader proxy, which renders the page and returns its text; and (6) an archived copy from the Internet Archive’s Wayback Machine. Reddit and Quora threads either returned only a search snippet or could not be adequately retrieved and were therefore excluded from comparison analyses.

\subsection{Analytic Sample}
Two gates defined the analytic sample (see Table~\ref{tab:app-sample}). Gate one was at the question level and required the AIOs own text to be present and at least 200 characters long; this removed seven overviews whose text capture failed. Gate two was at the page level and kept only readable pages. A question entered the paired comparisons when it had at least one readable AIO-cited page and at least one readable traditional result.

\begin{table}[tbp]
  \caption{Questions and Pages at Each Filter Stage}
  \label{tab:app-sample}
  \footnotesize
  \begin{tabular}{@{}p{3.3cm}rrr@{}}
    \toprule
    Stage & Questions & \shortstack[r]{AIO-cited\\pages} & \shortstack[r]{Traditional\\results} \\
    \midrule
    HealthBench questions submitted & 3,611 & -- & -- \\
    Search result captured & 3,610 & -- & -- \\
    AIO returned & 3,481 & 26,805 & 16,833 \\
    Gate 1: overview text of at least 200 characters & 3,474 & 26,780 & 16,803 \\
    Gate 2: readable pages only & -- & 23,657 & 14,443 \\
    At least one readable AIO-cited page & 3,137 & -- & -- \\
    At least one readable traditional result & 3,406 & -- & -- \\
    Paired analytic sample (both of the above) & 3,103 & 23,402 & 13,294 \\
    \bottomrule
  \end{tabular}
  \par\smallskip\raggedright \textit{Note.} Of the 129 captured searches without an AIO, 70 returned no overview and 59 returned an empty placeholder.
\end{table}

\begin{figure}[tbp]
  \centering
  \includegraphics[width=0.85\columnwidth]{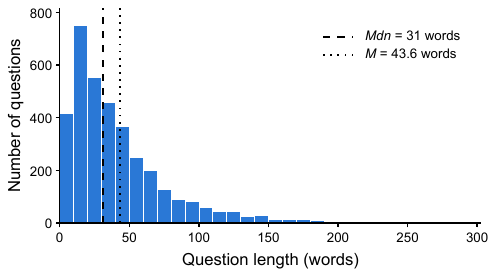}
  \caption{Distribution of HealthBench Question Length. \textit{N} = 3,611 questions. Bars are 10-word bins. The dashed line is the median (31 words) and the dotted line is the mean (43.6 words). Nine questions were 300 words or longer and are not shown.}
  \Description{Right-skewed histogram of question length in words, with most questions under 100 words.}
  \label{fig:app-length}
\end{figure}

\textit{Averaging Page Scores.} For each question and criterion, the readable AIO-cited pages and the readable traditional results were averaged separately, giving each question one mean per arm (for a single 0/1 criterion, the proportion of pages meeting it). Therefore, each question carries equal weight regardless of how many pages its overview cited.

\subsection{Page Signals Used for Coding}
If JAMA and HONcode criteria were judged manually, human coders would have judged a fully rendered page, using its bylines, “Medically reviewed by” lines, “Last updated” labels, footer disclaimers, and reference headings. Article-text extraction removes most of these, so coding from the extracted text alone would have missed them. The coding therefore used a second view of each HTML page, capturing its full visible text with only scripts and styles removed. Ten signals were taken from this view, and the JAMA and HONcode rules (Table~\ref{tab:app-jama}) are applied using these signals.

\textit{JAMA Benchmark Coding.} Authorship has one addition to the credential rule: a page on a trusted health domain passes even without a named credentialed author, because pages from these organizations are produced under known editorial responsibility. The trusted domains are the 476 organization domains linked from MedlinePlus, the National Library of Medicine’s consumer health site (see GitHub for the full list of approved domains).

\begin{table}[tbp]
  \caption{JAMA Benchmark Coding Rules}
  \label{tab:app-jama}
  \footnotesize
  \begin{tabular}{@{}p{1.55cm}p{6.25cm}@{}}
    \toprule
    Criterion & Scored 1 (passed) \\
    \midrule
    Authorship & A credentialed person is named as author, contributor, or medical reviewer, or the page is on a trusted health domain (MedlinePlus list, including subdomains). \\
    Attribution & The page has a reference-section heading, at least one numbered citation link, or at least two distinct external links in the article prose. \\
    Disclosure & The page carries a disclosure statement (a sentence of at least six words with a disclosure cue) or a named policy link. \\
    Currency & A date the reader can see: a labeled date anywhere on the page, an unlabeled date near the top of the article, or a publication date the scraper read from the visible page text. \\
    \bottomrule
  \end{tabular}
\end{table}

\textit{HONcode Coding.} The MedlinePlus trusted-domain rule applies only to JAMA Authorship; HONcode Authority uses its own list of institutional domains.

\textit{Agreement With Human Coders.} The rules were calibrated against a coding worksheet in which three coders coded the pages returned for three pilot questions on the original JAMA and HONcode criteria. Each of the 29 pages was coded by one coder, and every human judgment on the six retained criteria was compared with the automated code, giving 174 judgments (Table~\ref{tab:app-agree}).

\begin{table}[tbp]
  \caption{Agreement Between Automated and Human Coding}
  \label{tab:app-agree}
  \footnotesize
  \begin{tabular}{@{}lrrr@{}}
    \toprule
    Criterion & Judgments & Agreement (\%) & $\kappa$ \\
    \midrule
    JAMA Authorship & 29 & 90 & .75 \\
    JAMA Attribution & 29 & 79 & .56 \\
    JAMA Disclosure & 29 & 90 & .66 \\
    JAMA Currency & 29 & 83 & .66 \\
    HON Authority & 29 & 76 & .50 \\
    HON Complementarity & 29 & 86 & .26 \\
    \midrule
    All criteria & 174 & 83.9 & .68 \\
    \bottomrule
  \end{tabular}
  \par\smallskip\raggedright \textit{Note.} The low $\kappa$ for Complementarity despite 86\% agreement reflects that most pages carried a cue, which limits the agreement $\kappa$ can credit beyond chance.
\end{table}

\subsection{HallOumi Coding}\label{app:halloumi}
We used the classifier variant (oumi-ai/HallOumi-8B-classifier) loaded in 4-bit quantization on a single consumer GPU [NVIDIA RTX 2080 Ti, 11 GB] over the course of a week. Statements were enumerated at Google's native granularity: an overview is rendered as block units (i.e., a paragraph or a bullet) each carrying its own citation group, and each unit was treated as one statement that was never split across units. Each statement was decomposed into sentences, and sentences that are not verifiable claims (section lead-ins ending in a colon, navigational directives such as ``for more information, visit…'', bare URLs, and AI or medical disclaimers) were excluded from scoring. Because a cited page frequently exceeded the model's context window, each source was read in overlapping 1,600-token windows with 200-token overlap, so that no part of a page was discarded.

Scoring proceeded through two gates. At the sentence gate, a sentence counted as supported when any window of any readable source cited for that statement reached an entailment probability of 0.50. Sources cited elsewhere in the overview were not consulted, because the question of interest is whether Google's own citation supports the claim it is attached to. At the statement gate, a statement counted as supported when at least 50\% of its factual sentences passed the sentence gate.

To check HallOumi against human judgment, two coders each judged a different set of statements (60 and 40; 100 in total) as supported or not supported by the statement’s cited pages. Taking the coders as the reference, HallOumi’s sensitivity (statements the coders judged supported that HallOumi also judged supported) was 76.1\% (51 of 67) and its specificity (statements the coders judged not supported that HallOumi also judged not supported) was 90.9\% (30 of 33). We corrected the observed support rate with the Rogan–Gladen estimator~\cite{rogan1978}: corrected = (observed + specificity $-$ 1) / (sensitivity + specificity $-$ 1), which gives 56.7\% (95\% CI [45.3, 68.6]) for the observed 47.1\% (95\% CI [46.1, 48.2]).

\subsection{Physician Rubric Grading}
\begin{table*}[tbp]
  \caption{Main Results in the Full Sample and in Single-Turn Questions of 44 Words or Fewer}
  \label{tab:app-robust}
  \footnotesize
  \begin{tabular}{@{}p{5.6cm}p{5.4cm}p{5.4cm}@{}}
    \toprule
    Measure & Full sample & Single-turn, $\leq$ 44 words \\
    \midrule
    AI Overviews citing no sources (\%) & 9.4 [8.5, 10.4] & 5.8 [4.8, 7.1] \\
    \quad Low / medium / high complexity & 4.2 / 6.0 / 10.6; $\chi^2$(1) = 20.96*** & 3.4 / 4.0 / 6.5; $\chi^2$(1) = 4.27* \\
    \quad Low / medium / high risk & 6.4 / 8.1 / 17.6; $\chi^2$(1) = 55.57*** & 4.3 / 6.5 / 7.5; $\chi^2$(1) = 4.62* \\
    Source overlap (\%) & 44.4 [43.4, 45.4] & 46.7 [45.3, 48.2] \\
    \multicolumn{3}{@{}l}{Source quality, $d_z$ (AI Overview $-$ traditional)} \\
    \quad JAMA Attribution & $-$0.06 [$-$0.09, $-$0.02]** & 0.00 [$-$0.05, 0.05] \\
    \quad JAMA Disclosure & $-$0.12 [$-$0.16, $-$0.08]*** & $-$0.12 [$-$0.17, $-$0.07]*** \\
    \quad HON Authority & $-$0.19 [$-$0.22, $-$0.15]*** & $-$0.19 [$-$0.24, $-$0.14]*** \\
    \multicolumn{3}{@{}l}{Statement grounding (\%)} \\
    \quad Scorable statements supported & 47.1 [46.1, 48.2] & 49.9 [48.5, 51.5] \\
    \multicolumn{3}{@{}l}{Physician rubric scores (\% of available points)} \\
    \quad Clinical (accuracy and completeness) & 39.3 [38.0, 40.7] & 39.6 [37.5, 41.6] \\
    \multicolumn{3}{@{}l}{Clinical score and citations} \\
    \quad Spearman $\rho$ with JAMA total & $-$.05 [$-$.09, $-$.02]** & $-$.06 [$-$.11, $-$.01]* \\
    \quad Spearman $\rho$ with HONcode total & .10 [.06, .13]*** & .11 [.06, .16]*** \\
    \quad Spearman $\rho$ with grounding (support rate) & .01 [$-$.02, .05] & .06 [.01, .12]* \\
    \bottomrule
  \end{tabular}
  \par\smallskip\raggedright \textit{Note.} Brackets are 95\% confidence intervals: Wilson intervals for percentages of AI Overviews. Values separated by slashes are the low, medium, and high levels, followed by the Cochran–Armitage trend test (uncited overviews) or chi-square test (support). Negative $d_z$ means traditional results scored higher. *\textit{p} $<$ .05. **\textit{p} $<$ .01. ***\textit{p} $<$ .001; $d_z$ \textit{p} values are Benjamini–Hochberg adjusted across the six criteria.
\end{table*}

\textit{Rubrics.} The 41,180 rubric criteria for the 3,474 graded AIOs comprised 35,028 criteria written for the specific conversation and 6,152 consensus criteria shared across conversations; 28,962 described content to show and 12,218 content to avoid (median 12 per AIO, range 2–48).

\textit{Grader Validation.} HealthBench’s meta-evaluation file contains 29,511 model responses graded by physicians against its 34 consensus criteria (60,896 physician grades)~\cite{arora2025}. We drew 3,000 rows with the sampler in OpenAI’s simple-evals code (seed 0; 6,200 physician grades) and computed macro-F1 (MF1), the unweighted mean of the F1 scores for “met” and “not met,” between a grader and every physician grade of each row. GPT-4.1 configured from Arora et al.~\cite{arora2025} reached .707, in line with the .709 that Arora et al.~\cite{arora2025} reported.

\textit{Grading Procedure.} Each AIO was graded against each of its criteria in a separate request to GPT-4.1 (gpt-4.1) through OpenAI’s Batch API, with HealthBench’s grader prompt and settings unchanged~\citep[see][]{arora2025}. The conversation shown to the grader was the question as submitted to Google, with the user turns of the 1,445 multi-turn conversations joined by “ | ”, followed by the AIO text, with links and citation markers removed, as the response.

\textit{Scoring and Analysis.} Fourteen AIOs had no accuracy or completeness criteria and so were removed. Associations of clinical scores with the JAMA and HONcode totals and HallOumi support rates were tested with Spearman correlations on the 2,973 cited AIOs with all three measures. For the analysis without referral criteria, we removed HealthBench’s emergency-referral consensus criterion and every accuracy or completeness criterion about seeking or referring to care, identified by keywords (e.g., seek care, see a doctor, emergency department, call 911, referral, primary care, follow up with), from both the points earned and the points available. A minority of AIOs (\textit{N} = 247) left with no accuracy or completeness criteria were excluded.

\subsection{Known Limitations of the Automated Coding}
\textit{Text-only retrieval routes.} Pages delivered as plain text (through the Jina Reader proxy or the Europe PMC API, or as PDFs) carry no HTML structure. Therefore, reference headings, citation links, and prose links therefore could not be detected, and JAMA Attribution was always 0 for these pages. These routes supplied 15.7\% of readable AIO-cited pages and 16.2\% of readable traditional results in the paired sample, so the limitation lowers Attribution by a similar amount in both arms (i.e., rather than favoring one).

\textit{Unreadable pages.} Pages that could not be read (11.7\% of AIO-cited pages and 14.0\% of traditional results) were excluded rather than scored as 0, so sites that block automated access are under-represented in both arms.

\section{Additional Results}

\subsection{Analytic Sample}
We further characterized search overlap by examining whether the two arms used similar host sites (see Table~\ref{tab:app-domains}). Each domain was compared as a relative share of each arm using paired \textit{t}-tests (\textit{p} values Benjamini–Hochberg adjusted across the 12 domains).

\begin{table}[tbp]
  \caption{Most Frequent Domains Among AI Overview Citations and Traditional Results}
  \label{tab:app-domains}
  \footnotesize
  \setlength{\tabcolsep}{3pt}
  \begin{tabular}{@{}lrrrrr@{}}
    \toprule
    Domain & \shortstack[r]{AI Overview\\citations \textit{n} (\%)} & \shortstack[r]{Traditional\\results \textit{n} (\%)} & \textit{t}(3102) & \textit{p} & $d_z$ \\
    \midrule
    nih.gov & 2,965 (12.5\%) & 1,748 (12.1\%) & $-$1.46 & .172 & $-$0.03 \\
    clevelandclinic.org & 1,119 (4.7\%) & 1,158 (8.0\%) & $-$15.09 & $<$ .001 & $-$0.27 \\
    mayoclinic.org & 929 (3.9\%) & 839 (5.8\%) & $-$7.63 & $<$ .001 & $-$0.14 \\
    cdc.gov & 689 (2.9\%) & 495 (3.4\%) & $-$2.61 & .012 & $-$0.05 \\
    nhs.uk & 364 (1.5\%) & 442 (3.1\%) & $-$10.72 & $<$ .001 & $-$0.19 \\
    webmd.com & 471 (2.0\%) & 200 (1.4\%) & 5.48 & $<$ .001 & 0.10 \\
    healthline.com & 416 (1.8\%) & 181 (1.3\%) & 4.35 & $<$ .001 & 0.08 \\
    ubiehealth.com & 329 (1.4\%) & 129 (0.9\%) & 5.12 & $<$ .001 & 0.09 \\
    harvard.edu & 233 (1.0\%) & 175 (1.2\%) & $-$1.03 & .333 & $-$0.02 \\
    healthdirect.gov.au & 173 (0.7\%) & 185 (1.3\%) & $-$4.83 & $<$ .001 & $-$0.09 \\
    hopkinsmedicine.org & 173 (0.7\%) & 172 (1.2\%) & $-$4.39 & $<$ .001 & $-$0.08 \\
    medlineplus.gov & 186 (0.8\%) & 131 (0.9\%) & $-$0.45 & .655 & $-$0.01 \\
    \bottomrule
  \end{tabular}
  \par\smallskip\raggedright \textit{Note.} Negative \textit{t} and $d_z$ mean the domain made up a larger share of traditional results. Domains are registered domains, so nih.gov pools every NIH site (e.g., PubMed, PMC, and institute pages); ordered by combined count.
\end{table}

\subsection{Source Quality}\label{app:quality}
\begin{table}[tbp]
  \caption{JAMA and HONcode Scores of Pages Cited by AI Overviews and of the Traditional Top Five}
  \label{tab:app-quality}
  \footnotesize
  \setlength{\tabcolsep}{3pt}
  \begin{tabular}{@{}lrrrrrrr@{}}
    \toprule
    & \multicolumn{2}{c}{\shortstack{AI Overview\\citations}} & \multicolumn{2}{c}{Traditional top 5} & & & \\
    \cmidrule(lr){2-3}\cmidrule(lr){4-5}
    Measure & \textit{M} & \textit{SD} & \textit{M} & \textit{SD} & \textit{t}(3102) & \textit{p} & $d_z$ \\
    \midrule
    JAMA Authorship & 0.50 & 0.24 & 0.50 & 0.26 & $-$0.19 & .849 & 0.00 \\
    JAMA Attribution & 0.48 & 0.23 & 0.49 & 0.26 & $-$3.17 & .002 & $-$0.06 \\
    JAMA Disclosure & 0.32 & 0.22 & 0.35 & 0.24 & $-$6.68 & $<$ .001 & $-$0.12 \\
    JAMA Currency & 0.63 & 0.23 & 0.65 & 0.25 & $-$4.87 & $<$ .001 & $-$0.09 \\
    HON Authority & 0.59 & 0.24 & 0.64 & 0.26 & $-$10.45 & $<$ .001 & $-$0.19 \\
    HON Complementarity & 0.58 & 0.34 & 0.59 & 0.34 & $-$1.37 & .205 & $-$0.02 \\
    \bottomrule
  \end{tabular}
  \par\smallskip\raggedright \textit{Note.} Scores are per-question means over \textit{n} = 3,103 questions with at least one readable page on each side. Paired \textit{t}-tests; \textit{p} values are Benjamini–Hochberg (BH) adjusted across the six measures. $d_z$ = mean paired difference divided by the SD of the differences; negative \textit{t} and $d_z$ mean the traditional results scored higher.
\end{table}

\textit{Quality within Risk Level and Rubric Complexity.} To further explore differences in quality, we repeated the same paired comparison within each risk level and each rubric complexity level (\textit{p}-values were exploratory and therefore not corrected). HON Total favored traditional results at every risk level ($d_z$ from $-$0.19 to $-$0.10, all \textit{p} $<$ .001) and every complexity level ($d_z$ from $-$0.16 to $-$0.12, all \textit{p} $\leq$ .022). JAMA Total favored traditional results at every risk level ($d_z$ from $-$0.13 to $-$0.08, all \textit{p} $\leq$ .002) and every complexity level ($d_z$ from $-$0.14 to $-$0.10, all \textit{p} $\leq$ .050).

\textit{Physician Rubric Grading.} Table~\ref{tab:app-themes} reports clinical scores and the rate of content flagged as inaccurate or potentially harmful by HealthBench theme.

\begin{table}[tbp]
  \caption{Clinical Scores and Inaccurate or Harmful Content by HealthBench Theme}
  \label{tab:app-themes}
  \footnotesize
  \setlength{\tabcolsep}{3pt}
  \begin{tabular}{@{}llrrr@{}}
    \toprule
    Risk level & Theme & \textit{n} & \shortstack[r]{Clinical\\score (\%)} & \shortstack[r]{Inaccurate or\\harmful (\%)} \\
    \midrule
    Low & Health data tasks & 66 & 35.8 & 16.4 (55) \\
     & Context seeking & 495 & 35.2 & 14.3 (335) \\
     & Global health & 662 & 19.5 & 18.3 (410) \\
    Medium & Responding under uncertainty & 821 & 47.1 & 27.5 (625) \\
     & Expertise-tailored communication & 737 & 39.3 & 23.7 (333) \\
    High & Emergency referrals & 431 & 62.0 & 13.0 (146) \\
     & Response depth & 262 & 36.3 & 35.4 (161) \\
    \bottomrule
  \end{tabular}
  \par\smallskip\raggedright \textit{Note.} Clinical score = mean percentage of available points on accuracy and completeness criteria, clipped to 0–100\%. Inaccurate or harmful = percentage of AIOs meeting at least one negative-point accuracy criterion, among the AIOs (in parentheses) whose rubric contained one.
\end{table}

\subsection{Robustness to Question Length}
Table~\ref{tab:app-robust} repeats the main analyses on single-turn questions of 44 words or fewer (\textit{n} = 1,714 of 3,611 questions), the subset closest to typical search queries. The cutoff was chosen as the full-sample mean question length (43.6 words). Because the subset has fewer than half of the questions, estimates are compared by their direction and size (with 95\% confidence intervals) to the main results, rather than by statistical significance.

\end{document}